\documentclass[a4paper,11pt]{article}
\usepackage{theorem}

{\theorembodyfont{\rmfamily}
\newtheorem{definition}{Definition}[section]}
{\theorembodyfont{\rmfamily}
}
{\theorembodyfont{\rmfamily}
}

\usepackage{fullpage}
\begin{document}
\bibliographystyle{acm}
\frenchspacing
\title{Computing with P Systems\thanks{An slightly updated version of
A. Syropoulos, S. Doumanis, K.T. Sotiriades, Computing Recursive Functions with P Systems, 
Pre-proceedings of the Fifth Workshop on Membrane Computing (WMC5), Milano, Italy, June 2004, pp.~414--421.}}
\author{Apostolos Syropoulos,
        Stratos Doumanis
        and Konstantinos T. Sotiriades\\
        Greek Molecular Computing Group\\
                   Xanthi, Greece\\
        \texttt{asyropoulos@yahoo.com}}
\date{June 2004}
\maketitle
\begin{abstract}
P systems are computing conceptual computing devices that are at least
as powerful as Turing machines. However, until recently it was not known how 
one can encode any recursive function as a P~system. Here we propose a new
encoding of recursive as P~systems with graph-like structure, which is 
the main difference with previous documented attempts. The consequence
of this and other such efforts is that they provide a solid ground for
the implementation of real programming languages in existing hardware.
\end{abstract}
%%%%%%%%%%%%%%%%%%%%%%%%%%%%%%%%%%%%%%%%%%%%%%%%%%%%%%%%%%%%%%%%%%%%%
\section{Introduction}
%%%%%%%%%%%%%%%%%%%%%%%%%%%%%%%%%%%%%%%%%%%%%%%%%%%%%%%%%%%%%%%%%%%%%

Classically intractable and unsolvable problems are the driving force behind 
the ever increasing search for new computational models. These new models can 
be used to provide faster solutions to classically intractable 
problems~\cite{dna} and, in extreme cases, they can be {\em used} to 
{\em solve} classically unsolvable problems~\cite{quantum}. A major drawback
of such new models is that they seem to be useful only for the solution of a 
particular class of problems. For example, DNA computing has proved
efficient for the solution of combinatorial problems. Thus, a new computational
model is really useful once we have the techniques that allow us to compute
{\em ordinary} things. 

P systems are a new model of computation inspired by the way cells live and 
function (see~\cite{membrane} for an overview of the theory). Although, in
general, it has been proved that one can compute with P~systems whatever 
(universal) Turing machines are capable to compute; until recently it was not clear 
how one can compute primitive recursive functions with P~systems~\cite{pcomp2}.  

In this note we provide an alternative encoding of the basic functions as well as
three processes, which can be applied to the basic functions to yield
any $\mu$-recursive function. In order to encode these functions and processes
we had to introduce new rewriting rules and to make use of a more ``liberal''
membrane structures. Thus, our encoding has stretched to the limit the
capabilities of P~systems as far it regards numerical computation. 

In what follows, we present a very brief overview of the theory of P~systems. 
Next, we present our encoding and we conclude with some remarks concerning
our work and possible applications of it.

%%%%%%%%%%%%%%%%%%%%%%%%%%%%%%%%%%%%%%%%%%%%%%%%%%%%%%%%%%%%%%%%%%%%%
\section{Brief Overview of P Systems}
%%%%%%%%%%%%%%%%%%%%%%%%%%%%%%%%%%%%%%%%%%%%%%%%%%%%%%%%%%%%%%%%%%%%%
In what follows we will present an informal definition P~systems.
\begin{definition}
A P~system is an n-tuple 
\begin{displaymath}
\Pi=(O, \mu, w_1,\ldots, w_n, R_1,\ldots, R_m, i_0)
\end{displaymath}
where:
\begin{enumerate}
\item $O$ is an alphabet (i.e., a set of distinct entities) whose elements
      are called \textit{objects}.
\item $\mu$ is the membrane structure of the particular P~system; membranes
      are injectivelly labeled with succeeding natural numbers starting with
      one.
\item $w_i$, $1\le i\le m$, are strings that represent multisets over $O$
      associated with each region $i$.
\item $R_i$, $1\le i\le m$, are finite sets of rewriting rules (called 
      \textit{evolution rules}) over $O$. An evolution rule is of the form
      $u\rightarrow v$, $u\in O^{+}$ and $v\in O^{+}_{\mathrm{tar}}$,
      where $O_{\mathrm{tar}} = O\times\mathrm{TAR}$, 
      $\mathrm{TAR}=\{\mathrm{here},\mathrm{out}\}\cup\{\mathrm{in}_{j} |
      1\le j\le m\}$.
\item $i_0\in\{1,2,\ldots,m\}$ is the label of an elementary membrane (i.e.,
      a membrane that does not contain any other membrane), called the
      \textit{target} compartment. 
\end{enumerate}
\end{definition}
Note that the keywords ``here,'' ``out,'' and ``$\mathrm{in}_j$'' are called 
\textit{target} commands. Given a rule $u\rightarrow v$, the length of $u$ 
(denoted $|u|$) is called the radius of the rule. A P~system that contains 
rules of radius greater than one is a system with cooperation.

The rules that belong to some $R_i$ are applied to the objects of the
compartment $i$ synchronously, in a non-deterministic maximally parallel way.
Given a rule $u\rightarrow v$, the effect of applying this rule to a 
compartment $i$ is to remove remove the multiset of objects specified 
by $u$ and to insert the objects specified by $v$ in the regions designated 
by the target commands associated with the objects from $v$. In particular,
\begin{enumerate}
\item if $(a,\mathrm{here})\in v$, the object $a$ will be placed in 
       compartment $i$ (i.e., the compartment where the action takes
       place)
\item if $(a,\mathrm{out})\in v$, the object $a$ will be placed in 
       the compartment that surrounds $i$
\item if $(a,\mathrm{in}_{j})\in v$, the object $a$ will be placed in 
       compartment $j$, provided that $j$ is immediately inside $i$, or else
       the rule is not applicable
\end{enumerate}
%%%%%%%%%%%%%%%%%%%%%%%%%%%%%%%%%%%%%%%%%%%%%%%%%%%%%%%%%%%%%%%%%%%%%
\section{Encoding the Basic Functions as P Systems}
%%%%%%%%%%%%%%%%%%%%%%%%%%%%%%%%%%%%%%%%%%%%%%%%%%%%%%%%%%%%%%%%%%%%%
Our exposition on recursion theory is based on standard references~\cite{boolos,davis}.
Here are the three basic functions:
\begin{enumerate}
\item the {\em successor} function $S(x)=x+1$,
\item the {\em zero} function $z(x_1,\ldots,x_n)=0$, and
\item the {\em identity} function $U^{n}_{i}(x_1,\ldots,x_n)=x_i$,
      $1\le i\le n$.
\end{enumerate}
If we want to define $A$-recursive functions, we also need to use the characteristic
function. However, $A$-recursive sets are not important for our discussion, so we
are going to restrict ourselves to general recursive functions only. 

Two nested compartments (i.e., one inside another) make up the simplest
non-trivial membrane structure. For this reason, we have opted to encode
the basic functions as P~systems that have this particular
membrane structure. In what follows, the inner compartment will be the
target compartment. Let us now elaborate on the encoding of each function.

\paragraph{The zero function} This function simply discards its arguments
and returns the number zero. For each argument $x_i$ of the function we
pick up an object $\alpha_i$ and place $x_i$ copies of it in the outer
compartment. Next, we associate to each object $\alpha_i$ a multiset
rewrite rule of the form 
\begin{displaymath}
\alpha_i\rightarrow\varepsilon
\end{displaymath}
This is a new kind of rule that simply annihilates all objects as it replaces 
each object with the empty string. We can imagine that there is a pipe that
is used to throw the $\alpha_i$'s into the environment. This rule
can be considered to implement a form of ``catharsis'' of a compartment. Thus,
the outcome of the computation of this P~system is the number zero as
all objects are practically thrown to the environment. Obviously, this system 
implements the zero function.

\paragraph{The successor function} Suppose we have a P~system with no data
that, at the same time, encodes the successor function, then this system must 
place an object to the target compartment and stop. Obviously, the problem is
how this can happen, as there is no rule that can be applied to an empty 
compartment. For this reason, we introduce the new {\empty} rule
\begin{displaymath}
\varepsilon\rightarrow(\alpha, \mathrm{in}_{i})
\end{displaymath}
This rule can be applied only if there are no objects in a given compartment.
And its effect is to place an object $\alpha$ to compartment $i$. After this,
the rule is ``disassociated'' from the current compartment. In other words,
this rule can be used only once. The rationale behind the introduction of
this rule is that at any moment a cell can absorb matter from its
environment that, in turn, will be consumed. Having this new rule, it is
now trivial to encode the successor function. 

\paragraph{The identity function} This function is similar to the zero
function with one difference: it discards all of its arguments but one.
So, for each argument $x_i$ we place $x_i$ copies of an object $\alpha_i$.
Note that all objects $\alpha_i$ are distinct. Assume that the particular
instance of the identity function returns its $j$th argument, then we 
associate with $\alpha_j$ the following rule:
\begin{displaymath}
\alpha_{j}\rightarrow(\alpha_{j},\mathrm{in}_{2})
\end{displaymath}
Similarly to the zero function case, all other objects are associated
with a catharsis rule:
\begin{displaymath}
\alpha_{i}\rightarrow\varepsilon,\; i\not=j
\end{displaymath}
%%%%%%%%%%%%%%%%%%%%%%%%%%%%%%%%%%%%%%%%%%%%%%%%%%%%%%%%%%%%%%%%%%%%%
\section{Encoding the Processes as P Systems}
%%%%%%%%%%%%%%%%%%%%%%%%%%%%%%%%%%%%%%%%%%%%%%%%%%%%%%%%%%%%%%%%%%%%%
Recursive functions can be defined by applying function builders,
or processes, to the basic functions. For example, the addition function
is build by applying primitive recursion to the identity and successor
functions. Thus, it is necessary to provide an encoding of the three
processes in order to be able to compute any recursive function using
P~systems. The three processes are described below:
\begin{description}
\item[Composition] Suppose that $f$ is a function of $m$ arguments and
each of $g_1,\ldots,g_m$ is a function of $n$ arguments, then the function
obtained by composition from $f, g1, \ldots, g_m$ is the function $h$
defined as follows:
\begin{displaymath}
h(x_1,\ldots,x_n)=f(g_1(x_1,\ldots,x_n),\ldots,g_m(x_1,\ldots,x_n))
\end{displaymath}
\item[Primitive Recursion] A function $h$ of $k+1$ arguments is said to be
definable by (primitive) recursion from the functions $f$ and $g$, having 
$k$ and $k+2$ arguments, respectively, if it is defined as follows:
\begin{eqnarray*}
h(x_1,\ldots,x_k,0)&=&f(x_1,\ldots,x_k)\\
h(x_1,\ldots,x_k,S(m))&=&g(x_1,\ldots,x_k,m,h(x_1,\ldots,x_k,m))
\end{eqnarray*}
\item[Minimalization] The operation of minimalization associates with each
total function $f$ of $k+1$ arguments the function $h$ of $k$ arguments.
Given a tuple $(x_1,\ldots,x_k)$, the value of $h(x_1,\ldots,x_k)$ 
is the least value of $x_{k+1}$, if one such exists, 
for which $f(x_1,\ldots,x_k,x_{k+1})=0$.
If no such $x_{k+1}$ exists, then its value is undefined.
\end{description}

In the rest of this section we will not provide encodings for the general
case. Instead, we will concentrate on simple cases that can be easily
generalized. To put it in another way, if the P~systems that we present
faithfully encode the basic processes, then one can easily generalize
the encoding scheme. 

\paragraph{Encoding the composition process} 
Assume that $f$ is function that has two arguments
and that $g_1$ and $g_2$ are two functions with three arguments each, then
we want to design a P~system that will encode a function $h$ defined as
follows
\begin{equation}\label{eq:1}
h(x_1,x_2,x_3)=f(g_1(x_1,x_2,x_3),g_2(x_1,x_2,x_3))
\end{equation}
%%%%%%%%%%%%%%%%%%%%%%%%%%%%%%%B E G I N  F I G U R E %%%%%%%%%%%%%%%%%%%%%%%%
\begin{figure}
\begin{center}
\begin{picture}(200,200)
\put(0,0){\framebox(180,180)[cc]{}}
\put(5,170){$\alpha^{x_1}$}
\put(20,170){$\beta^{x_2}$}
\put(35,170){$\gamma^{x_3}$}
%%%%
\put(50,130){\oval(70,40)}
\put(65,130){\oval(20,10)}
\put(50,155){$g_1$}
%%%%
\put(130,130){\oval(70,40)}
\put(145,130){\oval(20,10)}
\put(130,155){$g_2$}
%%%%
\put(90,50){\oval(70,40)}
\put(90,75){$f$}
%%%%
\put(65,130){\line(1,-4){17}}
\put(67,130){\line(1,-4){17}}
%%%%
\put(145,130){\line(-1,-2){34}}
\put(147,130){\line(-1,-2){34}}
\end{picture}
\end{center}
\caption{A P system encoding the composition process.}
\end{figure}
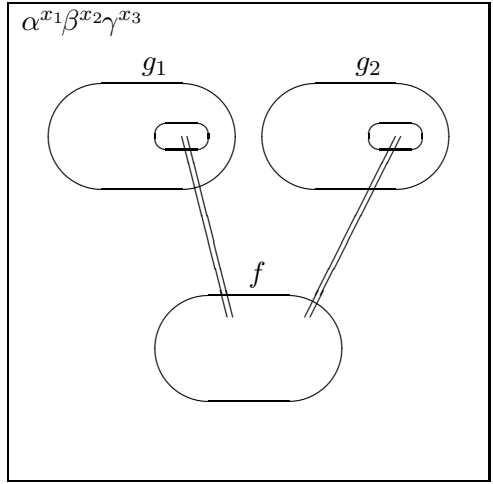
%%%%%%%%%%%%%%%%%%%%%%%%%%%%%%%END OF F I G U R E %%%%%%%%%%%%%%%%%%%%%%%%
Given three integers $x_1$, $x_2$ and $x_3$, we want to design a P~system
capable to compute $h(x_1,x_2,x_3)$. From equation~\ref{eq:1} it is clear
that we have first to compute $u_1=g_1(x_1,x_2,x_3)$ and 
$u_2=g_2(x_1,x_2,x_3)$. This suggests that the P~system will contain two
nested P~systems that will compute the numbers $u_1$ and $u_2$, respectively.
In addition, these numbers (i.e., a multiset of objects representing them)
have to be directly supplied to a third nested P~system, which will compute 
the value $u=f(u_1,u_2)$. This means that we need two pipelines that will
connect the output compartment of the first two nested P~systems with the
P~system that will compute the number $u$. Obviously, P~systems with
tree-like structure are not adequate to solve this problem. This means
that we need a way to more generally specify the communication channels
between the various compartments. An obvious solution, is to use P~systems
that have a (limited) graph-like structure. Indeed, in~\cite{pi-dist} the first
author of this paper provides such an extension.
\begin{definition}
A graph-structured symbol-object membrane system is a set of
(possibly nested) compartments that can be specified as follows:
\begin{displaymath}
\Pi=(O,g_m,w_1,\ldots,w_m,R_1,\ldots,R_m,i_0)
\end{displaymath}
Here:
\begin{enumerate}
\item $O$ is a set of objects that denote elements of information;
\item $g_m$ is a relation in $\{1,\ldots,m\}\times\{1,\ldots,m\}$,
      describing the network structure of the system;
\item $w_i$, $1\le i\le m$, are strings representing multisets of objects
      that are elements of $O$ that denote conglomerations of information;
\item $R_i$, $1\le i\le m$, are finite sets of \textbf{evolution rules} over
      $O$; $R_i$ is associated with compartment $i$; an evolution rule is
      of the form $u\rightarrow v$, where $u$ is a string over $O$ and $v$
      is a string over $O_{\mathrm{NTAR}}$, where $O_{\mathrm{NTAR}}=O\times
      \mathrm{NTAR}$, for $\mathrm{NTAR}=\{\mathrm{here}, \mathrm{out}\}
      \cup\{\mathrm{to}_{j}|1\le j\le m\}$;
\item $i_0 \in \{1,2,\ldots,m\}$ is the label of the \textbf{output
      compartment}.
\end{enumerate}
\end{definition}
The symbol ``$\mathrm{to}_j$'' should be used to directly place an object from
the host compartment to compartment $j$. The rule is applicable only if
$(i,j)\in g_m$ or $(j,i)\in g_m$.

\begin{figure}
\begin{center}
\begin{picture}(300,200)
\put(0,0){\framebox(270,190)[cc]{}}
\put(5,180){$\alpha^{x}$}
\put(20,180){$\beta^{y}$}
%%%%
\put(20,100){\framebox(50,30){}}
\put(40,110){\framebox(20,10){}}
\put(30,140){$f(x,y)$}
\put(58,117){\line(1,0){40}}
\put(58,114){\line(1,0){40}}
\put(90,100){\framebox(50,30){}}
\put(110,110){\framebox(20,10){}}
\put(90,140){$g(x,0,f(x))$}
\put(128,117){\line(1,0){40}}
\put(128,114){\line(1,0){40}}
\put(172,115){\ldots}
\put(190,117){\line(1,0){20}}
\put(190,114){\line(1,0){20}}
\put(200,100){\framebox(50,30){}}
\put(220,110){\framebox(20,10){}}
\put(210,140){$g(\ldots)$}
\put(230,115){\line(0,-1){60}}
\put(233,115){\line(0,-1){60}}
\put(200,30){\framebox(50,30){}}
\end{picture}
\end{center}
\caption{A P system encoding primitive recursion.}\label{fig:2}
\end{figure}
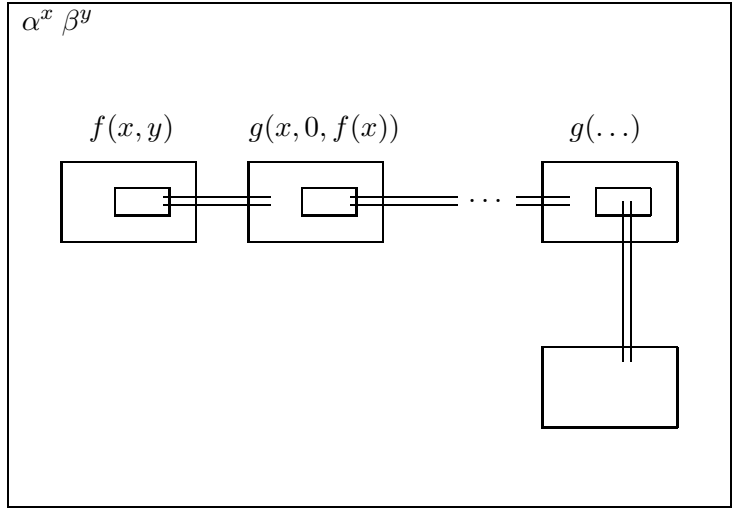

%%%%%%%%%%%%%%%%%%%%%%%%%%%%%%%%%%%%%%%%%%%%%%%%%
\paragraph{Encoding primitive recursion}  
%%%%%%%%%%%%%%%%%%%%%%%%%%%%%%%%%%%%%%%%%%%%%%%%%
In primitive recursion we a priori know the number of iterations. Thus, if we want to
compute $h(x,y+1)$ we actually have to compute the following:
\begin{displaymath}
g(x,y,g(x,y-1,g(x,y-2,\ldots g(x,0,f(x))\ldots)))
\end{displaymath}
Thus, if we want to compute $h(x,y)$, we need $n$ compartments that will
compute the successive values of $g$ and one compartment that will compute the value of
$f(x)$. In addition, all these compartments must form a pipeline where the $i$th
compartment will get its data from compartment $i-1$ and it will feed its output to
the compartment $i+1$. Naturally, the last compartment will populate with data the
output compartment. Initially, all compartments will get the same copies of a
designating object, say \#, which will denote the first argument of function $h$. Also,
the compartments that compute the successive values of $g$ will get the respective number
of copies of another designated object, say @. Figure~\ref{fig:2} depicts the initial
setting of the system. 
%%%%%%%%%%%%%%%%%%%%%%%%%%%%%%%B E G I N  F I G U R E %%%%%%%%%%%%%%%%%%%%%%%%
%%%%%%%%%%%%%%%%%%%%%%%%%%%%%%%END OF F I G U R E %%%%%%%%%%%%%%%%%%%%%%%%

\begin{figure}
\begin{center}
\begin{picture}(200,200)
\put(0,0){\framebox(200,170)[cc]{}}
\put(5,160){$\alpha^{x}$}
\put(20,160){$\beta^{y}$}
\put(5,5){$1$}
%%%%
\put(30,100){\framebox(50,30){}}
\put(50,110){\framebox(20,10){}}
\put(40,140){$f(x,y)$}
\put(66,117){\line(1,0){40}}
\put(66,114){\line(1,0){40}}
\put(33,103){$2$}
\put(100,100){\framebox(50,30){}}
%\put(120,110){\framebox(20,10){}}
\put(103,103){$3$}
\put(130,110){\line(0,-1){60}}
\put(133,110){\line(0,-1){60}}
\put(100,30){\framebox(50,30){}}
\put(103,33){$4$}
\end{picture}
\end{center}
\caption{A P system encoding minimalization function.}\label{fig:3}
\end{figure}
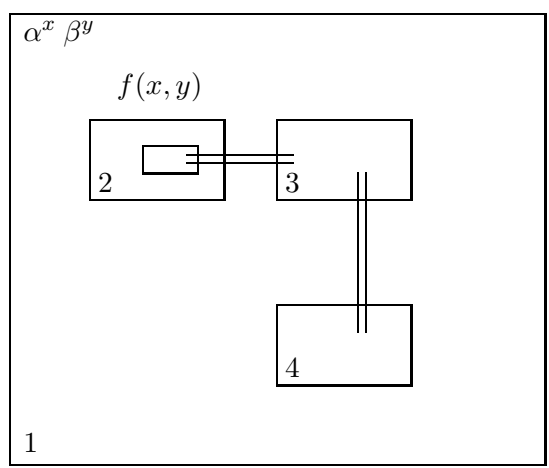

%%%%%%%%%%%%%%%%%%%%%%%%%%%%%%%%%%%%%%%%%%%%%%%%%%%%%%%%%%
\paragraph{\textbf{Encoding the minimalization function builder}}
%%%%%%%%%%%%%%%%%%%%%%%%%%%%%%%%%%%%%%%%%%%%%%%%%%%%%%%%%%
As for the previous cases, we present a simplified system that computes function
$h(x)$. Assume that $f$ is a function of two arguments $x$ and %y$. Our primary target is 
to compute $f(x,y)$ with a fixed $x$, which is equal to the argument of the original
function $h$, and a variable $y$, which is increased by one in each iteration, that halts 
the computation the very moment the outcome of the function computation is zero. 
Figure~\ref{fig:3} shows the general setup of a P~system that computes function $h(x)$. 
Compartment~3 is directly connected to the output compartment of compartment~2, where 
actual the computation is taking place. The output of this compartment is pipelined to
compartment~3 and it is transformed to the distinguished object \#. This symbol is
used as counter. The connection from the compartment
number~3 to compartment number~4 is used to move the object \# to it. In addition, at
each iteration, the $a$'s and $b$'s are also placed in compartment~3 in the form of
of two distinguished elements $\oplus$ and $\otimes$. More specifically, we move the
$a$'s and the $b$'s in compartment~1 and at the same time we create a number of
$\oplus$'s and $\otimes$'s that is to number of $a$'s and $b$'s respectively. Clearly,
if the result of the computation at compartment~2 is the number zero, no action should
take place at compartment~3. Otherwise, we first move \# to compartment~4 and place
a copy of $b$ at the compartment~1, then we 
place the $\oplus$'s and $\otimes$ to compartment~1 in the form of $a$'s and $b$'s.
This way, we actually increase the $b$'s by one, which is what we actually need to
do to get the correct result.

%%%%%%%%%%%%%%%%%%%%%%%%%%%%%%%B E G I N  F I G U R E %%%%%%%%%%%%%%%%%%%%%%%%
%%%%%%%%%%%%%%%%%%%%%%%%%%%%%%%END OF F I G U R E %%%%%%%%%%%%%%%%%%%%%%%%
It is important to note that in all cases our description is a little bit vague,
as we do not give the full details of the encoding. However, our intension was
to give the general idea of how things can be done. After all, we believe it is not
difficult to work out these details so to provide a complete description of the
encoding of each function builder.
%%%%%%%%%%%%%%%%%%%%%%%%%%%%%%%%%%%%%%%%%%%%%%%%%%%%%%%%%%%%%%%%%
\section{Conclusions}
%%%%%%%%%%%%%%%%%%%%%%%%%%%%%%%%%%%%%%%%%%%%%%%%%%%%%%%%%%%%%%%%%
We have presented an alternative encoding of the basic functions and the three
function builders of recursion theory. Since any functional programming language
can be used to compute any recursive function, one can build a front-end capable
of compiling (simple) functional programs into P~systems. Indeed, in~\cite{pi-dist},
the first author of this paper presents ideas related to this particular implementation
procedure. We believe that more work is needed to be done in order to provide 
a complete and viable solution to the problem of encoding recursive functions as
P~systems. We just hope that our work is step towards this direction.
%%%%%%%%%%%%%%%%%%%%%%%%%%%%%%%%%%%%%%%%%%%%%%%%%%%%%%%%%%%%%%%%%
%\bibliography{Pcomp}
%%%%%%%%%%%%%%%%%%%%%%%%%%%%%%%%%%%%%%%%%%%%%%%%%%%%%%%%%%%%%%%%%

\end{document}